\documentclass[useAMS,usenatbib]{mn2e}
\usepackage{epsfig,amssymb}
\title[SED and HE emission of BL Lacs]{
Spectral energy distributions and high-energy emission of BL Lac type
objects } \author[S.~Troitsky]{Sergey Troitsky\thanks{E-mail:
st@ms2.inr.ac.ru}\\
Institute for Nuclear Research of the Russian Academy of
Sciences,
60th October Anniversary prospect 7a, 117312, Moscow, Russia
}
\begin{document}
\date{}
\pagerange{\pageref{firstpage}--\pageref{lastpage}} \pubyear{2008}
\maketitle
\label{firstpage}
\begin{abstract}
Based on identifications from the V\'eron and Quasars.org catalogs, we
determine the optical-to-X-ray spectral indices for a sample of 201 BL
Lac type objects (BLLs) and search for trends in the distribution of these
indices of the sources detected in high-energy bands.
We find that EGRET-detected sources are low-energy peaked and that the
positional correlation with the arrival directions of ultra-high-energy
cosmic rays from the previously studied AGASA, Yakutsk and High Resolution
Fly's Eye samples is dominated by low-energy-peaked BLLs.
\end{abstract}


\section{Introduction}
\label{sec:intro}
BL Lac type objects (hereafter BLLs) attract considerable attention of both
astrophysicists and particle physicists. Among other reasons, it is
caused by observation of very-high-energy ($\sim 1$~TeV) gamma rays from
these sources (see e.g.\ \citet{TeV-rev,wagner} and references therein) and
by claims for potential association of BLLs with ultra-high-energy
(UHE; $\gtrsim 10^{19}$~eV) cosmic rays (CRs)
\citep{TT:BL,TT:BL-GMF,GTTT:EGRET,GTTT:HiRes,HiRes:BL}. Most probably,
only particular subclasses of BLLs are responsible for
these high-energy emissions. More precise determination of these
subclasses may shed light on the intrinsic mechanisms of particle
acceleration in blazars and may help in determination of potential sources
yet unobserved at high or ultra-high energies.

Broadband spectral energy distributions (SEDs) of BLLs are known to have
a very specific non-thermal two-bump shape. In a
popular model, the two bumps are caused by the synchrotron and
inverse-Compton radiations (see e.g.\ \citet{2bump1,2bump2}). Their origin
is attributed to other mechanisms in some models
(see e.g.\ \citet{Mannheim}). While the overall two-bump shape is quite
common, the location of the peaks varies strongly from source to source.
In particular, the well-measured low-energy peak corresponds to infrared
or optical frequencies in the so-called low-energy-peaked BLLs (LBLs)
and to X-ray frequencies in the high-energy-peaked objects (HBLs). As it
has been recently understood, intermediate cases are also present. The
correlation of the peak position with the intrinsic power of the object
\citep{blazar-seq} is currently under discussion (see e.g.\
\citet{blazar-seq-discuss}). The second, often worse measured peak is
located in the gamma-ray band (MeV to GeV in LBLs and hundreds of GeV in
HBLs).

Since LBLs are bright in the optical band and relatively faint in X rays
(keV frequencies correspond to a dip between two peaks in this case) while
HBLs peak in the X-ray band, the optical-to-X-ray broadband spectral index
$\alpha_{\rm OX}$ provides a good measure of the position of the first
peak in SED (see e.g.\ \citet{Donato}). Detailed measurements of the SED
are available for a small fraction of BLLs while to know $\alpha_{\rm
OX}$ one needs only optical and X-ray measurements performed for a much
wider sample of sources. In this note, we use $\alpha_{\rm OX}$ as
the quantity which characterizes the SED of a BLL; results of a more
detailed multiwavelength study will be reported elsewhere. We
determine $\alpha_{\rm OX}$ for a large sample of BLLs and look
for a correlation between the index and the high-energy emissivity of a
source. We will see that
EGRET-detected sources are mostly LBLs.
One of the primary goals of this study is to
specify the class of BLLs which correlate with the arrival directions
of ultra-high-energy cosmic rays; we will see that the correlations
observed previously are also saturated by LBLs.

The rest of the Letter is organized as follows. In Sec.~\ref{sec:sample},
the sample of BLLs is discussed and the index $\alpha_{\rm OX}$ is
determined. In Sec.~\ref{sec:gamma}, we demonstrate the trends in the
distribution of $\alpha_{\rm OX}$ of EGRET sources. In
Sec.~\ref{sec:uhecr}, we briefly review previously reported UHECR -- BL
Lac correlation and discuss the distribution of $\alpha_{\rm OX}$ of
correlated objects. Brief conclusions are summarized in
Sec.~\ref{sec:concl}.

\section{The sample}
\label{sec:sample}
We study objects classified as confirmed BLLs (class BL or
HP) in \citet{Veron} (hereafter the V\'eron catalog). Objects from this
catalog were searched for X-ray identifications in ROSAT data in the
Quasars.org catalog \citep{qorg}.
There, the V\'eron objects have been cross-correlated with various
published ROSAT catalogs (\citet{1RXSbright,1RXSfaint,2RXP,1RXH,WGA1}) and
the probability of the true identification was presented for each case. Of
the sources from different catalogs associated with a given BLL, we select
one with the highest probability;
we drop the source from the sample if this probability is
below 68\%. In such a way, we
obtain a sample of 201 objects with $V$-band magnitudes given in
the V\'eron catalog and X-ray identifications with one of the ROSAT
catalogs. We note that in some cases, the name of the object in the
V\'eron catalog has the ROSAT prefix but this identification has very low
probability according to Quasars.org and the corresponding BLL was
therefore dropped from the sample.

For the objects identified with the ROSAT latest PSPC catalogs
\citep{1RXSbright,1RXSfaint,2RXP}, we calculate the flux in $(0.1 \div
2.4)$~keV band from the count rate and the hardness ratio following
\citet{1RXSbright}. In a few cases when the best identification
is with the catalog of \citet{WGA, WGA1}, we take the flux value presented
there. If the object was observed by the high-resolution imager (HRI), the
best identification is often with the catalog of \citet{1RXH}; we use the
flux values given in the BMW catalog \citep{BMW} in these cases. These
latter fluxes correspond to the same energy band and assume a Crab-like
spectrum for each source. It is a relatively rough approximation caused by
a poor spectral sensitivity of HRI; however, for firm identifications, the
BMW and PSPC fluxes are well correlated.

For each object in the sample, we calculate the optical-to-X-ray spectral
index $\alpha_{\rm OX}$ using the textbook definition
\citep{CarrollOstlie},
$$
\frac{F_{\rm O}}{F_{\rm X}}=\left(\frac{\nu _{\rm O}}{\nu_{\rm X}} \right)
^{-\alpha_{\rm OX}},
$$
where $F_i\, d\nu _i$ is the amount of energy with frequencies
between $\nu _i$ and $\nu _i+d\nu _i$ per unit area per second, observed
by a detector aimed at the source and measured e.g.\ in W/(cm$^2\cdot$s);
index $i$ stands here for either $V$-band, $i=$O, or X-ray band, $i=$X,
frequencies and fluxes.

Most of the BLLs are strongly variable at all wavelengths and therefore
the use of non-simultaneous observations may introduce random errors
as large as 0.3 to $\alpha_{\rm OX}$. Given the fact that the X-ray
sample is essentially flux-limited, these errors may be asymmetric for
faint objects.

An important correction to $V$ and therefore to $\alpha_{\rm OX}$ may
arise from the contribution of the host galaxy emission. We use the
magnitudes from the V\'eron catalog which are not corrected for the
contribution of the host galaxies. These corrections might cause
significant systematic errors in $\alpha_{\rm OX}$ given the fact that the
host galaxy contribution is more important in the visible light than in X
rays. The effect of this correction is stronger in fainter, less beamed or
misaligned BLLs. Account of this correction on the case-by-case basis
would result in a significant reduction of the statistics since for many
objects in the sample (notably for those associated with EGRET sources or
correlated with UHECR) the host galaxy was not detected, indicating either
powerful or well beamed source. Another approach is to use infrared colour
indices to estimate, on a statistical basis, whether the contribution of a
host galaxy is important. Developed by \citet{Glass}, this approach was
systematically applied to V\'eron BLLs by \citet{Chen}, where $J$, $H$ and
$K$ magnitudes were obtained from the 2MASS survey and the objects for
which a significant contribution of the host galaxy is expected were
determined. To estimate the effect of the contamination by the emission of
the host galaxies on our results, we removed these objects (37 out of 201)
from our sample and checked that the effects we advocate are present in
the reduced sample as well\footnote{Of our sample, 34 objects are not
included in the catalog of \citet{Chen}. We assume that the lack of
infrared identification implies negligible contribution of the host
galaxy for these sources.}, see below. Since the starlight of the host
galaxy is expected to peak in the infrared (see e.g.\ discussion in
\citet{Kotilainen}), the effect in the $V$-band should be even smaller.

Our sample is derived from the V\'eron catalog which is incomplete.
Therefore, the results of the study may be biased; it is not guaranteed
that they are generic and hold for all BLLs in the Universe. However, by
comparing different subsets of one and the same catalog, we partially
remove the effects of the incompletness and reveal important trends which
can be tested in future studies. Though the selection effects may be
different for different subsamples, the most important of them are under
control in the present study.

\section{Gamma-ray emitters}
\label{sec:gamma}
The identification of BLL counterparts of the EGRET sources is a
nontrivial task and becomes sometimes a subject of debates. There exist
EGRET sources identified with BLLs by one authors but identified with
other objects (e.g.\ clusters of galaxies) by others. For this study, we
select all potential BLL identifications from
\citet{3EG,Mattox,GTTT:EGRET} and divided them into ``high-probability''
and ``others'' following the identification probabilities from
\citet{Mattox} (the high-probability subsample includes objects from
Tables~1 and 2 of \citet{Mattox}).

\begin{figure}
\begin{center}
\includegraphics[width=0.95 \columnwidth]{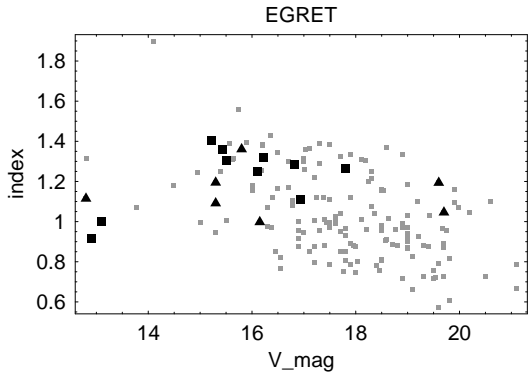}
\caption{
Optical-to-X-ray spectral index $\alpha_{\rm OX}$ versus $V$-band
magnitude for all BLLs in the sample (gray) and for EGRET-detected
objects (black). Black boxes correspond to probable identifications
(according to \citet{Mattox}), triangles correspond to less probable or
not studied by \citet{Mattox} objects.
\label{fig:EGRET-mag-ind}
}
\end{center}
\end{figure}

\begin{figure}
\begin{center}
\includegraphics[width=0.95 \columnwidth]{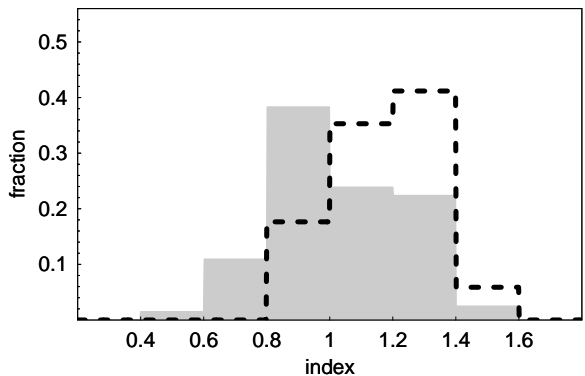}
\caption{
Distribution of $\alpha_{\rm OX}$ for all BLLs in the sample (gray) and
for objects possibly detected by EGRET (dashed).
\label{fig:EGRET-ind}
}
\end{center}
\end{figure}

In Fig.~\ref{fig:EGRET-mag-ind},
we present magnitudes and spectral indices of those BLLs in our
sample which are associated with catalogued EGRET sources.
Fig.~\ref{fig:EGRET-ind} compares the distribution of the spectral indices
of correlated objects with that of all objects in the sample. In agreement
with the two-bump synchrotron--self--Compton SED model, EGRET-selected
sources are low-energy peaked. To quantify this statement, we compared the
distribution of $\alpha_{\rm OX}$ for the EGRET BLLs and that for all
objects in the sample by means of the Kolmogorov-Smirnov (KS) test. The
probability that the parent distribution is the same is $P\approx 5 \cdot
10^{-3}$; it reduces to $\approx 10^{-2}$ if only high-probability
identifications are used or if the BLLs with potentially strong
contribution of the host galaxy are dropped. The change in the KS
probability corresponds to the decrease of the sample size. We note that
all 14 high-probability EGRET identifications correspond to BLLs without
significant expected contribution from the host galaxy, so a more detailed
account of this bias, would it be possible, may only strengthen our
conclusion.

A similar study for the BLLs detected at very high
energies ($E \gtrsim 200$~GeV) is hardly possible because they\footnote{
The list of these sources may be found on the MAGIC
webpage
\tt
http://www.mppmu.mpg.de/$\tilde{~}$rwagner/sources/index.html
}
are not drawn from a full-sky survey but from pointed observations of
X-ray selected objects (low $\alpha_{\rm OX}$).
One should note however that strong X-ray flux does not guarantee that a
BLL is a TeV emitter; indeed MAGIC performed a dedicated search for TeV
emission from X-ray selected blazars and in many cases did not discover it
\citep{MAGIC:survey}.
We note in passing that no significant
preference in $\alpha_{\rm OX}$ is found for the BLLs which are possible
emitters of 10-GeV gamma-rays \citep{GTTT:MNRAS}; they have, on average,
intermediate values of $\alpha_{\rm OX}$ between EGRET and TeV emitters.

\section{Cosmic-ray correlation}
\label{sec:uhecr}
Previous studies reported significant correlation between various samples
of BLLs from the V\'eron catalog and various samples of UHECRs. Some of
the studies \citep{TT:BL-GMF,GTTT:EGRET} used reconstruction of the
arrival directions in the Galactic magnetic field assuming charged
cosmic-ray particles; we will not discuss them here because of the
ambiguity in the magnetic-field models (see e.g.\ \citet{Cao:GMF}). On the
other hand, a number of studies suggest correlation which assumes zero
deflection (neutral particles); these results are less model-dependent and
much more intriguing because neutral UHE particles from BLLs
would challenge conventional models of cosmic-ray physics. These claims
include the correlation found in the sample of cosmic rays observed by the
Akeno Giant Air-Shower Array (AGASA; sample with estimated primary
energies $E>4.8 \cdot 10^{19}$~eV) and the Yakutsk Extensive Air Shower
Array (Yakutsk; $E>2.4 \cdot 10^{19}$~eV) detectors where an excess of
pairs `BLL -- cosmic ray' was seen at separations less than 2.5$^\circ$
\citep{TT:BL} and in a sample of events with $E>10^{19}$~eV observed by
the High Resolution Fly's Eye detector (HiRes) for separations less than
0.8$^\circ$ \citep{GTTT:HiRes}. In both cases the separation was
consistent with the detector's angular resolution (which was much better
in HiRes than in AGASA and Yakutsk). The correlation with the HiRes sample
was confirmed in an unbinned study  and was found to be held at lower
energies using unpublished data \citep{HiRes:BL}. The probability to
observe the correlation with three independent experiments by chance was
estimated by \citet{comparative} as $3\cdot 10^{-5}$ by a Monte-Carlo
study which took into account statistical penalty for multiple tries
(various catalogs of potential sources tested for correlation).

\begin{figure}
\begin{center}
\includegraphics[width=0.95 \columnwidth]{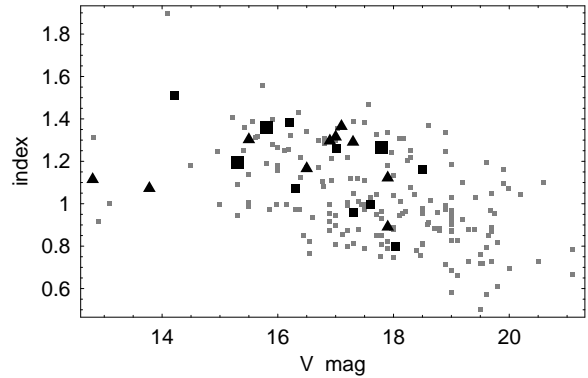}
\caption{
Optical-to-X-ray spectral index $\alpha_{\rm OX}$ versus $V$-band
magnitude for all BLLs in the sample (gray) and for those correlated
with arrival directions of ultra-high-energy cosmic rays (black).
Black boxes denote objects correlated with AGASA and Yakutsk cosmic rays
(large boxes correspond to objects correlated with doublets),
triangles denote objects correlated with HiRes cosmic rays (see text for
more details).
\label{fig:all-mag-ind}
}
\end{center}
\end{figure}

\begin{figure}
\begin{center}
\includegraphics[width=0.95 \columnwidth]{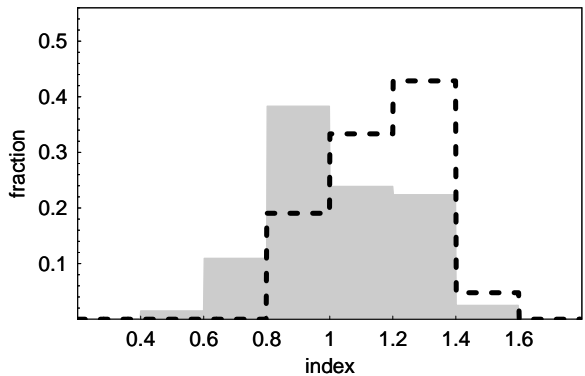}
\caption{
Distribution of $\alpha_{\rm OX}$ for all BLLs in the sample (gray) and
for objects correlated with ultra-high-energy cosmic rays (dashed).
\label{fig:all-ind}
}
\end{center}
\end{figure}

In Fig.~\ref{fig:all-mag-ind},
we present magnitudes and spectral indices of those BLLs in our
sample which are located within $2.5^\circ$ of the arrival directions of
AGASA and Yakutsk cosmic rays of the sample used in \citet{TT:BL} and
within $0.8^\circ$ of those of the HiRes sample used in
\citet{GTTT:HiRes}. Fig.~\ref{fig:all-ind} compares the distribution of
the spectral indices of correlated objects with that of all objects in
the sample. We see that the correlation is dominated by low-energy peaked
BLLs; the KS test for the two distributions gives
$P\approx 4 \cdot 10^{-3}$.
This confirms and improves our preliminary approximate results
\citep{Vietnam} based on a sample of BLLs with X-ray identifications
`by eye'. The results of the present study (cf.\ Fig.~\ref{fig:EGRET-ind}
and Fig.~\ref{fig:all-ind}) support also the association between UHECR and
EGRET sources suggested by \citet{GTTT:EGRET}. This association is not
surprising since both acceleration and propagation of UHECR are inevitably
accompanied by emission of secondary energetic photons which in turn
interact with the cosmic background radiation and lose their energy until
it reaches $\sim 0.1 \div 10$~GeV.

The discussion of the potential bias due to host galaxies is very similar
to the case of the EGRET sources. Restriction of the sample to objects
with no expected host galaxy contribution results in the KS probability of
$P\approx 10^{-2}$, change being consistent with the decrease of the
sample size. Of 21 objects in the sample associated with UHECRs, only 3
may be affected by the host galaxy, according to \citet{Chen}. This fact
is in agreement with previous observations \citep{TT:BL} that the UHECR
correlation is stronger for objects with unknown redshifts.

The correlation between BLLs and UHECRs seen in HiRes data
\citep{GTTT:HiRes} has been tested recently by the Pierre Auger (PA)
Collaboration \citep{Auger:BL}; no positive signal was found. This is not
conclusive however for the following reasons. Firstly, PA is located in
the Southern hemisphere and sees different BLLs than other experiments;
moreover, due to incompleteness of the catalogs, the number of
potential UHECR emitters is much less in the South. For instance, of 99
objects in our sample which have $\alpha_{\rm OX}>1$ (which seem to
correlate with UHECR stronger, see Fig.~\ref{fig:all-ind}), 82 can be seen
by HiRes but only 46 are in the field of view of PA, most of them only
at large zenith angles and for a small fraction of time.
The `factor of merit' depends also on the angular resolution of the
experiment which is twice worse in the PA surface detector than in HiRes
(stereoscopic mode).  This problem was quantified by \citet{predictions}
where it has been shown that to reach the HiRes sensitivity for the signal
found by \citet{GTTT:HiRes} in the set of 271 events, PA has to
accumulate $\sim 3500$ events in the same energy range. We note that
\citet{Auger:BL} used 1672 events for the test of this signal.

Secondly, as it has been pointed out by \citet{GTTT:HiRes} and
\citet{HiRes:BL} and further discussed by \citet{TT:neutral}, the
correlation observed by HiRes implies neutral cosmic particles
travelling for cosmological distances, the fact which requires
unconventional physics. Most probably the primary particles of the
resulting air showers are neither protons, nor nuclei. However, the
energy determination of the PA surface detector is extremely sensitive to
the type of the primary cosmic particle because of very strong sensitivity
of water tanks to muons in the air shower. For instance, energies of gamma
rays are always underestimated by a factor of a few (see e.g.\
\citet{Billoir}). Due to the steeply falling spectrum of UHECRs, this may
dilute the observed signal. PA possesses fluorescent detectors similar to
those of HiRes, but the BLL correlation was never tested with them.
Future tests of the correlation should be performed with high-resolution
fluorescent detectors, preferably in the Northern hemisphere, the
Telescope Array (see e.g.\ \citet{TA}) providing a good example.

\section{Conclusions}
\label{sec:concl}

We have studied the distibution of broadband optical-to-X-ray spectral
indices $\alpha_{\rm OX}$ of 201 confirmed BLLs from the V\'eron
catalog which have at least 68\% confident X-ray identification in the
Quasars.org catalog. In accordance with the
synchrotron--self--Compton two-bump SED models, a subsample of
EGRET-detected BLLs has on average high $\alpha_{\rm OX}$ (sources are
strong in optical but relatively faint in X rays).
One of the most important results of
the study relates to the BLLs correlated with ultra-high-energy cosmic
rays: most of them are low-energy peaked.
Both for the EGRET and UHECR associations, the contribution of the host
galaxy is expected to be negligible according to the infrared photometry
of \citet{Chen}. This fact agrees well with the expectations that the
EGRET sample is dominated by beamed sources and that the UHECR sources are
powerful and possess strongly collimated jets pointing precisely to the
observer. The present study narrows the class of potential UHECR emiters
to those with  high $\alpha_{\rm OX}$.
This fact may be used for
further tests of BLL -- cosmic ray correlation with fluorescent detectors
and may shed light on the origin of the correlated cosmic-ray particles. If
confirmed, this correlation would point to completely new phenomena in
particle physics or/and astrophysics because no known neutral particles of
these energies are expected to travel for distances larger than $\sim
10$~Mpc.

The author is indebted to the anonymous referee for numerous interesting
and valuable comments and to P.~Tinyakov for careful reading of the
manuscript and useful discussions. I thank D.~Bergman, D.~Gorbunov,
B.~Komberg, J.~Matthews, G.~Rubtsov, L.~Scott, G.~Thomson and I.~Tkachev
for interesting discussions. This work was supported in part by the grants
NS-7293.2006.2 (government contract 02.445.11.7370) and RFBR 07-02-00820.


\bsp

\label{lastpage}

\end{document}